\documentclass[a4paper]{article}
\pdfoutput=1
\usepackage[english]{babel}
\usepackage[utf8]{inputenc}
\usepackage{natbib}
\usepackage{amsmath}
\usepackage{wasysym}
\usepackage{graphicx}
\usepackage{url}
\usepackage{listings}
\usepackage{array}
\usepackage{verbatim}
\usepackage{longtable}
\usepackage[official]{eurosym}
\usepackage[ruled,linesnumbered]{algorithm2e}
\usepackage{tikz}
\usepackage{booktabs}
\usetikzlibrary{shapes,arrows, shapes.geometric, shapes.symbols}
\everymath{\displaystyle}
\usepackage{authblk}

\tikzstyle{block} = [rectangle, draw, fill=blue!20, 
    text width=3cm, text centered, rounded corners, minimum height=4em]
\tikzstyle{line} =  [draw, thick, ->, shorten >=2pt] 
\tikzstyle{cloud} = [draw, ellipse,fill=red!20, node distance=3cm,
    minimum height=2em]
\tikzstyle{print} = [draw, tape, tape bend top=none, fill=blue!20, node distance=3cm
	 text width=5em, minimum height=4em, text centered]
\tikzstyle{decision}= [diamond, aspect=2, draw, fill=blue!20,
                        text width=2.5cm, text centered,
                        inner sep=1pt, rounded corners]

\title{Empowering cash managers to achieve cost savings by improving predictive accuracy}

\author[1]{Francisco Salas-Molina\footnote{Corresponding author. E-mail addresses: \textit{francisco.salas@hifesa.com, martin@bigml.com, jar@iiia.csic.es, joan.serra@telefonica.com, arcos@iiia.csic.es}}}
\author[2]{Francisco J. Martin}
\author[3]{Juan A. Rodr\'iguez-Aguilar}
\author[4]{Joan Serr\`a}
\author[3]{Josep Ll. Arcos}
\affil[1]{Hilaturas Ferre, S.A., Les Molines, 2, 03450 Banyeres de Mariola, Alicante, Spain}
\affil[2]{BigML, Inc, 2851 NW 9th Suite, Conifer Plaza Building, Corvallis, OR 97330, US}
\affil[3]{IIIA-CSIC, Campus UAB, 08913 Cerdanyola, Catalonia, Spain}
\affil[4]{Telefonica Research, Ernest Lluch i Martín, 5, 08019 Barcelona, Catalonia, Spain}

\begin{document}
\bibliographystyle{apa}
\setcitestyle{authoryear,open={(},close={)}}
\maketitle

\begin{abstract}
Cash management is concerned with optimizing the short-term funding requirements of a company. To this end, different optimization strategies have been proposed to minimize costs using daily cash flow forecasts as the main input to the models. However, the effect of the accuracy of such forecasts on cash management policies has not been studied. In this article, using two real data sets from the textile industry, we show that predictive accuracy is highly correlated with cost savings when using daily forecasts in cash management models. A new method is proposed to help cash managers estimate if efforts in improving predictive accuracy are proportionally rewarded by cost savings. Our results imply the need for an analysis of potential cost savings derived from improving predictive accuracy. From that, the search for better forecasting models is in place to improve cash management.
\end{abstract}

\section{Introduction}

Cash flow management is concerned with the efficient use of a company's cash as a critical task in working capital management. Decision making in cash flow management focuses on keeping the balance between what the company holds in cash and what has been placed in short-term investments, such as deposit accounts or treasury bills. Different models have been designed to answer to these questions and reviews can be found in \cite{gregory1976cash,srinivasan1986deterministic,da2015stochastic}. However, to the best of our knowledge, little attention has been placed on the utility of cash flow forecasts with the exception of \citet{stone1972use} and \citet{gormley2007utility}. Both works showed the utility of forecasting in cash management, but none of them researched the importance of the predictive accuracy of the forecasts used in their respective models. Therefore, it is unknown whether even small improvements in predictive accuracy may lead to savings that could perhaps amount to millions of euros in total. 

The corporate cash management problem was first addressed from an inventory control point of view by \citet{baumol1952transactions} in a deterministic way. \citet{miller1966model} introduced a simple stochastic approach by considering a symmetric Bernouilli process in which both the inflow and the outflow were exactly of the same size and had probability $1/2$. Later, while \citet{girgis1968optimal} considered continuous net cash flows with both fixed and linear transaction costs, \citet{eppen1969cash} focused on discrete net cash flows with only variable transaction costs. The use of forecasts in the corporate cash management problem was first introduced by \citet{stone1972use} as a way of smoothing cash flows. More recently, \citet{gormley2007utility} used the model proposed by \citet{penttinen1991myopic} as a benchmark to demonstrate the utility of cash flow forecasts in the cash management problem. They proposed a dynamic simple policy to minimize transaction costs, under a general cost structure, and developed a time series model to provide forecasts. Surprisingly, even though their model is based on cash flow forecasts, they ignored the possibility of exploring alternative forecasting methods. We claim this step as a mandatory one, specially when improving forecasting accuracy may be correlated with cost savings. This hypothesis was suggested by the authors but was not verified. 

In cash flow forecasting, \citet{stone1976payments,stone1981daily,stone1987daily}, \citet{stone1977daily}, \citet{miller1985daily}, and \citet{maier1981short} presented different useful linear models. A measure of quality of any forecasting technique is its predictive accuracy. However, under an economic perspective, predictive accuracy must be mapped to estimated cost savings. This analysis allows to know how much companies can save by improving predictive models and, consequently, the cost of not predicting, i.e., the missed savings minus the cost of implementing the model. For example, if a reduction of 32\% in forecasting error produced \euro320.000 in savings per year, it can be stated that, on average, each percentage point of predictive accuracy is \euro10.000 worth.

Overall, one can conclude that when using forecasts in cash management models, predictive accuracy of these forecasts attracted little attention of the research community, neglecting its significance and implications. Consequently, our discussion leads to assess the quality of alternative forecasting methods. In this paper, we present and compare different forecasting methods including linear and non-linear models. In this sense, we expect that non-linear models such as radial basis functions and random forests can deal with cash flow time-series in a cost-saving approach. 

Using two real data sets from companies in the textile industry in Spain, in this paper:

\begin{itemize}
\item We show empirically that forecasting accuracy is highly correlated with savings in cash management. Thus, a comparison in terms of accuracy and savings between different forecasting models is performed. 
\item We argue that the effect of forecasting accuracy on cash management can be estimated in advance. Thus, a new methodology for estimating this effect is proposed.
\end{itemize}

The rest of this paper is organized as follows. We firstly describe our real cash flow data sets in Section \ref{section:data}. We later enumerate different forecasting models: linear models such as autoregressive and regression models; and non-linear models, such as radial basis function models and random forests in Section \ref{section:models}. These forecasting models will be ranked according to its predictive accuracy in the evaluation Section \ref{section:Evaluation}. In Section \ref{section:PredAcComp}, we empirically verify that a better forecast produces a better policy. Moreover, we estimate how much savings (if any) can be obtained by the cash policies produced by an improvement in forecasting accuracy. Finally, Section \ref{sec:conclusion} concludes.

\section{Description and data preprocessing \label{section:data}}

\begin{figure}[htb]
\centering
\includegraphics[width=1\textwidth,height=0.45\textwidth]{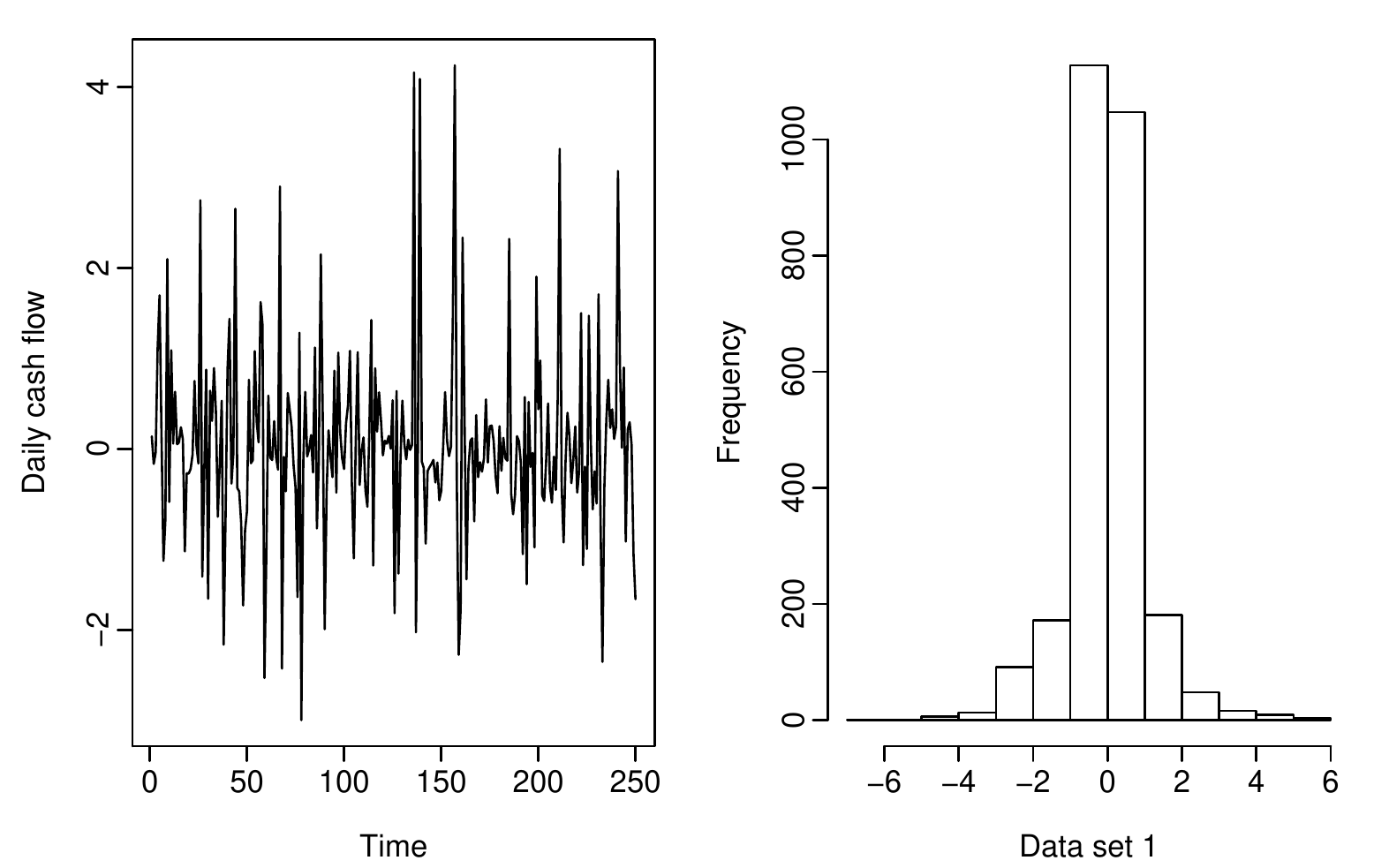}
\caption{\label{fig:DCF}Normalized net daily cash flow sample plot and histogram for data set 1 (adimensional).}
\end{figure}

In this section, we describe the two real cash flow data sets used in this paper. Data sets 1 and 2 gather net daily flow on workdays from two different companies in the textile industry. Both sets of observations are in the domain of real numbers and their values' distributions present a bell-like shape but excess kurtosis. A sample of our real net daily cash flow is shown in Figure \ref{fig:DCF}. For confidentiality reasons, figures show demeaned data divided by the standard deviation. Besides, an additional transformation is performed to deal with anomalies. More specifically, following the recommendations in \citet{gormley2007utility}, any observation greater than five times the standard deviation is replaced by a value of magnitude exactly five times the standard deviation. 

In order to cover a wider range of realistic industrial company cases, a number of cash flow data sets are derived from the two real data sets as follows:

\begin{itemize}
\item Real cash flow: data sets 1 and 2.
\item Stable cash flow: data set 3 is derived from data set 1 and applies to companies in a more stable environment with daily cash flows characterized by a low variance. In this case, observations greater than three times the standard deviation are replaced by values of exactly three times the standard deviation.
\item Unstable cash flow: data set 4 also derives from data set 1 and applies for companies with high variance in its daily cash flow due to different reasons such as a reduced number of customers or suppliers as it is the case of small companies. In this case, observations greater than three times the standard deviation are replaced by values of exactly two times the original observation.
\item Random shock cash flow: data sets 5 and 6 are derived from data sets 1 and 2 respectively and aim to cover the likely occurrence of unexpected changes in industrial markets. In this case, 5\% of the observations are randomly chosen and replaced by values drawn from a uniform distribution between the minimum and the maximum of the original time series.
\end{itemize}

A summary of the characteristics of each data set is presented in Table \ref{tab:DataSum}.

\begin{table}[h]
\caption{Data set summary.}
\begin{center}
\begin{tabular}{ c r l r r }
\hline
Data Set & Length & Case & Std.Dev. & Kurtosis  \\ \hline
1    & 2717         & Real cash flow  & 95745         & 4.82  \\ 
2    & 1218         & Real cash flow  & 44733         & 5.19 \\
3    & 2717         & Stable cash flow    & 89467         & 2.51  \\
4    & 2717         & Unstable cash flow   & 132170       & 20.66  \\
5    & 2717         & Random shock cash flow      & 113208       & 4.58    \\
6	 & 1218         & Random shock cash flow      & 54270         & 4.85	\\
\hline
\end{tabular}
\end{center}

\label{tab:DataSum}
\end{table}

For comparison purposes with \citet{gormley2007utility}, here we assume that, apart from daily cash flow data, no other extra features are provided by the company. However, a set of available explanatory variables can be proposed. According to \citet{miller1985daily}, and \citet{stone1977daily}, we may find seasonal patterns in daily cash flow data. Thus, we consider basic calendar effects such as the day-of-week effect and the day-of-month effect by using categorical or dummy variables. In the latter case, each dummy variable takes a value of one if time $t$ occurs at the corresponding day of the week/month, and zero otherwise. Month and week dummy variables are also added to the set of the explanatory variables. A further step in the search of explanatory power is explored by considering past values of the time series. From that, a tentative set of explanatory variables is listed below:  

\begin{itemize}
\item Day-of-month: Day of month categorical variables or dummy variables ($d_{t1},\ldots, d_{t31}$).
\item Day-of-week: Day of week categorical variables or dummy variables for working days ($s_{t1},\ldots,s_{t5}$).
\item Month: Month dummy variables ($m_{t1}, \ldots, m_{t12}$).
\item Week: Week dummy variables ($w_{t1}, \ldots, w_{t53}$).
\item Past values: Previous observations of the daily cash flow time series ($y_{t-1}, \ldots,y_{t-p}$) where $p$ is the total number of observations considered.
\end{itemize}

From a combination of these explanatory variables, different predictive models can be built and compared in terms of forecasting accuracy.

\section{Building forecasting models \label{section:models}}

The accuracy of any forecasting model depends on its ability to capture the specific characteristics of the data used. According to \citet{stone1972use}, real-world cash flows are neither completely known in advance nor are they completely unpredictable. However, a wide set of tools and techniques are available to improve forecasting accuracy. We claim that exploring alternative models to improve forecasting ability is mandatory, specially if improving forecasting accuracy can lead to cost savings. 

In this section, we present a number of forecasting models to be evaluated allowing us to identify our best-in-class forecaster that will ultimately be used as the main input to the cash management model. We do not intend to determine the best cash flow forecaster among all methods presented in forecasting research literature. Instead, our final goal is to verify if a better forecaster, in terms of forecasting accuracy, is able to produce a better cash policy in terms of cost savings. In this sense, we expect that non-linear models outperform two of the most usual linear models in cash flow forecasting allowing cash managers to deploy better cash policies.

Then, we here consider four different forecasting models: autoregressive, regression, radial basis functions and random forests. Firstly, we follow the autoregressive approach to daily cash flow forecasting along the lines of \citet{gormley2007utility}. In contrast to such approach, we expect that the use of the set of explanatory variables mentioned earlier rather than only a number of previous values of a time series can help obtain a more accurate prediction. Then, we secondly consider a linear regression model with a set of explanatory variables. 

While linear models are often employed in finance due to their simplicity, many non-linear models have been proposed to explain financial phenomena. Perhaps one of the most widely known non-linear model in finance is the \citet{black1973pricing} option pricing model. Moreover, there is a reason for optimism about the use of non-linear models in time-series prediction and finance as stated in \citet{weigend1994time,kantz2004nonlinear,small2005applied}. Firstly, several limitations of linear models were pointed out by \citet{miller1985daily} in daily cash flow forecasting such as interactions and holiday effects. Additionally, statistical hypothesis such as normality and stationarity are required by linear models to produce reliable results. On the other hand, alternative approaches to discover the relationship between time-series observations are also available. In this sense, non-linear models allow to explore beyond the constraints imposed by linear models through a much wider class of functions. 

Although non-linear time series analysis is not as well established as its linear counterpart \citep{de200625}, works by \citet{terasvirta2006forecasting, bradley2004forecasting,clements2004forecasting,sarantis2001nonlinearities,conejo2005forecasting} constitute good examples of its application to finance and economics. In this paper, we consider two non-linear models such as radial basis functions and random forest models due to the lack of attention of the research community. Next, we briefly describe our selection of forecasters and provide details on the implementation of non-linear models.

\subsection{Autoregressive model\label{subsec:ar}}

A widespread linear model in time series data is the autoregressive (AR) process, where predictions are based on a linear combination of previous observations \citep{box1976time}. AR models for cash flow forecasting can be found in \citet{gormley2007utility,laukaitis2008functional}. On the other hand, as mentioned in Section \ref{section:data}, our cash flow data have a bell-like shape but excess kurtosis. Hence, we follow the recommendations in \citet{gormley2007utility} and use an extension of the Box-Cox transformation described in \citet{box1964analysis} to approximate our data to a Gaussian distribution by tuning a parameter $\lambda$. Predictions are assessed using the following equation:

\begin{equation}
y_t^{(\lambda)}=\beta_0+\sum_{i=1}^{p}\beta_iy^{(\lambda)}_{t-i}+\epsilon
\label{eq:AR}
\end{equation}
where $y^{(\lambda)}$ is the cash flow forecast at time t, $[y_{t-1}^{(\lambda)},y_{t-2}^{(\lambda)}\cdots,y_{t-p}^{(\lambda)}]$ are the \emph{p}-previous observations of a transformed time series, $\beta_i$ is the \emph{i}-th estimation coefficient, and $\epsilon$ stands for the prediction error. Superscript $(\lambda)$ in both forecasts and previous observations denotes data transformation. This transformation is reversible and, therefore, $y_t$ can be derived from $y_t^{(\lambda)}$. When training the model, the number of previous observations, \emph{p}, is automatically chosen by minimizing the Akaike Information Criteria (AIC) using the \emph{ar} function in R \citep{R} for the autoregressive model. AIC is a measure of the quality of a time series model that is usually accepted as a selection criterion \cite{akaike1974new}.

\subsection{Regression model}

An autoregressive model is only based on the previous observations of the time series and misses possible patterns, if any, hidden in the data. When dealing with daily data, these patterns refer to calendar variations such as holidays, day of the month or day of the week. Trying to identify these patterns, here we consider a general regression model based on different explanatory variables. Regression models have been used for cash flow forecasting purposes in \citet{stone1977daily,stone1987daily,miller1985daily}. In this case, it is important to say that the ability of the modeler in the search for the best explanatory variables plays a key role. A general regression model is represented by the following equation:

\begin{equation}
y_t=\sum_{i=0}^{n}\beta_ix_{ti}+\epsilon.
\label{eq:reg}
\end{equation}

In this general model (\ref{eq:reg}) we relate $y_t$, the value of the cash flow at time \emph{t} to a linear combination of explanatory variables $x_{t1},x_{t2},\ldots,x_{tn}$ at the same time \emph{t}, being $\beta_i$ the \emph{i}-th regression coefficient, and $\epsilon$ the prediction error. From the general model (\ref{eq:reg}), a number of particular models can be derived for predictions depending on the different explanatory variables considered. For the implementation of these models we use the \emph{lm} function in R.

\subsection{Radial basis function model}

Financial data are usually originated by complex systems that may include non-linear processes. In order to capture non-linearity in the data, we also consider Radial Basis Function (RBF) models as described in \citet{weigend1994time,broomhead1988multivariable}. To use an RBF model, we first partition the input space by applying the $k$-medoids algorithm \citep{park2009simple} over the training set. Then a scalar Gaussian RBF $\phi(x)$ is used for forecasting:

\begin{equation}
y_t=b_0+\sum_{k=1}^{K}b_k\phi(\|x_{t}-c_k\|)+\epsilon
\label{eq:rbf}
\end{equation}
where $y_t$ is the value of the target variable at time \emph{t}, \emph{K} is the total number of clusters, $b_k$ is the coefficient associated to the \emph{k}-th cluster, $c_k$ is the \emph{k}-th cluster medoid, $x_{t}$ is the input data point at time \emph{t}, $\|\,\,\|$ is the Euclidean distance and $\epsilon$ is the prediction error. Finally, $\phi(x)$ is the following Gaussian function:
\begin{equation}
\phi(x,\alpha)=\mathrm{e}^{-x^2/\alpha\rho_k}
\end{equation}
where $\alpha$ is a positive integer parameter and $\rho_k$ is the mean distance between the elements inside the \emph{k}-th cluster. In this case, predictions are produced using our tentative set of explanatory variables and general matrix functions in R. 

Next, we provide an example of predictions obtained using RBF for the last 3 days of data set 1 based on the previous 21 cash flow observations. Two parameters have to be chosen to produce forecasts using RBF: the total number of clusters $K$, defining the degree of partition of the input space and the parameter $\alpha$, determining the contribution of deviate points to the prediction. For the sole purpose of this example, we set $K=5$ and $\alpha=10$. Then, we proceed as follows:
 
\begin{enumerate}
\item We create the input space $(2714-21)\times21$ matrix $X$ by embedding in each row 21 consecutive cash flows. Firstly, we firstly transform cash flows to $y_t^{(\lambda)}$, as explained in Section \ref{subsec:ar}. To avoid high values bias, we later standardize transformed cash flows by demeaning and dividing by the standard deviation.
\item We create a column vector $\textbf{y}$ of length 2693 with subsequent cash flows. 
\item We select each cluster medoids $c_k$ from rows in $X$ using the k-medoids algorithm.
\item We compute $\rho_k$ as the mean Euclidean distance between the elements of the $k$-th cluster to its medoid $c_k$.
\item We compute the $2693 \times 5$ matrix $\Phi$ where each row contains distances computed using the function $\phi(\|x_{t}-c_k\|)$ for each point in the input space to each cluster.
\item We obtain the column vector $\textbf{b}$ of weights by solving $\textbf{b}=(\Phi^T\Phi)^{-1}\Phi^T \textbf{y}$ using least squares. 
\item We produce a $3\times 21$ matrix $\hat{X}$ with the previous 21 observations prior to each of the 3 cash flows to be predicted and a $3\times 6$ matrix $\hat{\Phi}$ where the first column is set to 1 and the rest of elements are distances computed using the function $\phi(\|x_{t}-c_k\|)$ for each point in $\hat{X}$ to each cluster medoid.
\item We forecast by means of $\hat{y}=\hat{\Phi}\textbf{b}$, that has to be re-scaled by multiplying by the standard deviation and adding the mean and, finally, $\lambda$-transformed.
\end{enumerate}

Now, we are in a position to compare these forecasts to real values and to other forecasts and compute predictive accuracy as we will see below.

\subsection{Random forest model}

Decision trees are non-linear models that split the input space in subsets based on the value of a particular feature. On the other hand, an ensemble methodology is able to construct a predictive model by integrating multiple trees in what is called a decision forest \citep{dietterich2000ensemble}. Regression forests are used for the non-linear regression of dependent variables given independent inputs based on an ensemble of slightly different trees. Particularly, random forests (RF) are ensembles of randomly trained decision trees \citep{ho1995random,ho1998random,criminisi2013decision}. Recent examples of time series forecasting using random forests can be found in \citet{kumar2006forecasting,kane2014comparison,mei2014random,zagorecki2015prediction}.

We make predictions using the R package \emph{randomForest} by \cite{liaw2002random} which implements Breiman's random forest algorithm for classification and regression \citep{breiman2001random}. In this paper, we limit ourselves to select three parameters: the number ($a$) of randomly trained trees, the number ($b$) of variables randomly sampled as candidates at each split, and the node size ($c$) or the minimum amount of observations in a terminal node used to control overfitting. 

For instance, assume we know there is a strong daily seasonality in our cash flow. One possible way to assess how strong is this seasonality is to produce predictions using two explanatory variables: Day-of-month and Day-of-week. Hence, we aim to create a random forest model and predict the last 100 days of data set 1 based on these two variables. An example on how to proceed is as follows:

\begin{enumerate}
\item Create a $2617\times2$ matrix $X$ containing in each row the Day-of-month and the Day-of-week of past cash flows. 
\item Create a column vector $\textbf{y}$ of length 2617 with the corresponding cash flows.
\item Create a model based on $X$ and $\textbf{y}$, with $a=100$ randomly trained trees, with $b=2$ randomly sampled variables and node size $c=50$. 
\item Produce a $100\times2$ matrix $\hat{X}$ with the Day-of-month and the Day-of-week of last 100 cash flows of data set 1.
\item Input matrix $\hat{X}$ to the model to obtain the forecasts.
\end{enumerate}

Now, we are in a position to assess the importance of each explanatory variable or to test the quality of our predictions as we do next.

\section{Forecasting models' comparison \label{section:Evaluation}}

In this section, we aim to evaluate the forecasting accuracy of the presented models for comparison purposes. Autoregressive, regression, radial basis function and random forests models may produce different predictions with different accuracy. The comparison will allow us to determine our best-in-class forecaster to be later used as the input to establish the best cash management policy available. More precisely, we use time series cross-validation for different prediction horizons (\emph{h}) from 1 up to 100 days ahead by comparing the mean square error $\varepsilon(h)$ for different models:

\begin{equation}
\varepsilon(h)=\frac{\sum_{test}(\hat{y}_{t+h}-y_{t+h})^2}{\sum_{test}(\overline{y}-y_{t+h})^2}
\label{eq:MSE}
\end{equation}
where $h$ is the prediction horizon in days, $\hat{y}_{t+h}$ is the prediction at time ${t+h}$, $y_{t+h}$ is the real observation at the time ${t+h}$, and $\overline{y}$ is the the arithmetic mean of the real observations on the training set. Note that the closer $\varepsilon$ is to zero, the better the predictive accuracy. If $\varepsilon$ is close to one, the performance is similar to that of the mean as a naive forecast. Values greater than one show that the forecaster has no predictive ability. 

In \citet{hyndman2013forecasting}, two different time series cross-validation approaches were suggested: one with a fixed origin for the training set, and one with a rolling origin. Algorithm \ref{GeneralAlg} implements these two cross-validation methods.
\\

\begin{algorithm}[H]
 \textbf{Input:} Cash flow data set of \emph{T} observations, \emph{FixedOrigin}, minimum number \emph{g} of observations to forecast and prediction horizon \emph{h}\;  
 \textbf{Output:} Forecast accuracy for different prediction horizons\;
 \For{$i=1,2,\ldots,T-g-h+1$}
  {
  Select the observation at time $g+h+i-1$ for the test set\;
  \eIf{$FixedOrigin=True$}
  	{Estimate the model with observations at times $1,2,\ldots,g+i-1$\;}   
  	{Estimate the model with observations at times $i,i+1,\ldots,g+i-1$\;}
  Compute the \emph{h}-step error on the forecast for time $g+h+i-1$\;
  }
 Compute $\varepsilon(h)$ based on the errors obtained\;
 \caption{Time series cross validation algorithm}
 \label{GeneralAlg}
 \end{algorithm}
\vspace{5mm}

If the binary variable \emph{FixedOrigin} is set to True, the training set is formed by all the observations that occurred prior to the first observation that forms the test set (Method 1). We can get rid of the oldest observations by setting \emph{FixedOrigin} to False (Method 2) and considering only the \emph{g} most recent values (e.g., the last two or three years) by applying a sliding window of observations. In both methods we assume that a minimum number of \emph{g} observations is required to produce a reliable forecast. In our experiments, high values of \emph{g} produced almost no difference between Method 1 and Method 2. Using Method 2, smaller values of \emph{g} in steps of 250, equivalent to 1 year of observations, were also tried with worse results. Because of that, here we only present results for Method 1.

When parameters selection was necessary, an evaluation of the coefficient of determination, $R^2$, over a training set with the oldest 65\% of the observations was performed to choose the best value for each of the parameters. An exploration of the tentative explanatory variables mentioned above was performed. More precisely, we computed the error $\varepsilon(h)$ for $h \in[1,2,\ldots,100]$ using Algorithm \ref{GeneralAlg} (Method 1) for different tentative forecasting models with data sets 1 and 2. The first 65\% of the data was considered as the minimum length of the data to train the model. Results showed that the day-of-month and the day-of-week presented the best forecasting ability using RF models with data set 1 (Table \ref{tab:compCF1a}) and using RBF with data set 2 (Table \ref{tab:compCF2a}). 

\begin{table}[h]
\caption{Model characterization and average error ratio ($\overline{\varepsilon}$) for horizons up to 100 days (Data set 1).}
\begin{center}
\begin{tabular}{l l l l}
\hline
Model & Input variables & Parameters & $\overline{\varepsilon}$ (Std.Dev.) \\ \hline 
AR & $p$ past values & $p$ coefficients & 1.00 (0.007)  \\ 
REG & $d_{t2}, \ldots, d_{t31},s_{t1},\ldots, s_{t,4}$ & 35 coefficients & 0.70 (0.007)  \\ 
RBF & 20 past values, Day-of-month & $K=35, \alpha=10$ & 0.88 (0.003) \\
RF & $d_{t2}, \ldots, d_{t31},s_{t1},\ldots, s_{t4}$ & $a=20, b=11, c=50$ & 0.68 (0.010) \\
\hline
\end{tabular}
\end{center}
\label{tab:compCF1a}
\end{table}

\begin{table}[h]
\caption{Model characterization and average error ratio ($\overline{\varepsilon}$) for horizons up to 100 days (Data set 2).}
\begin{center}
\begin{tabular}{l l l l} 
\hline
Model & Input variables & Parameters & $\overline{\varepsilon}$ \space(Std.Dev.) \\ \hline 
AR & $p$ past values & $p$ coefficients & 1.00  (0.002)  \\
REG & $d_{t2}, \ldots, d_{t31},s_{t1},\ldots, s_{t4}$ & 35 coefficients & 0.96 (0.008)  \\ 
RBF & Day-of-month, day-of-week & $K=10, \alpha=10$ & 0.93 (0.008) \\
RF & Day-of-month, day-of-week & $a=20, b=11, c=50$ & 0.94 (0.006)\\
\hline
\end{tabular}
\end{center}
\label{tab:compCF2a}
\end{table} 

\begin{figure}[!htb]
\centering
\includegraphics[width=0.8\textwidth,height=0.5\textwidth]{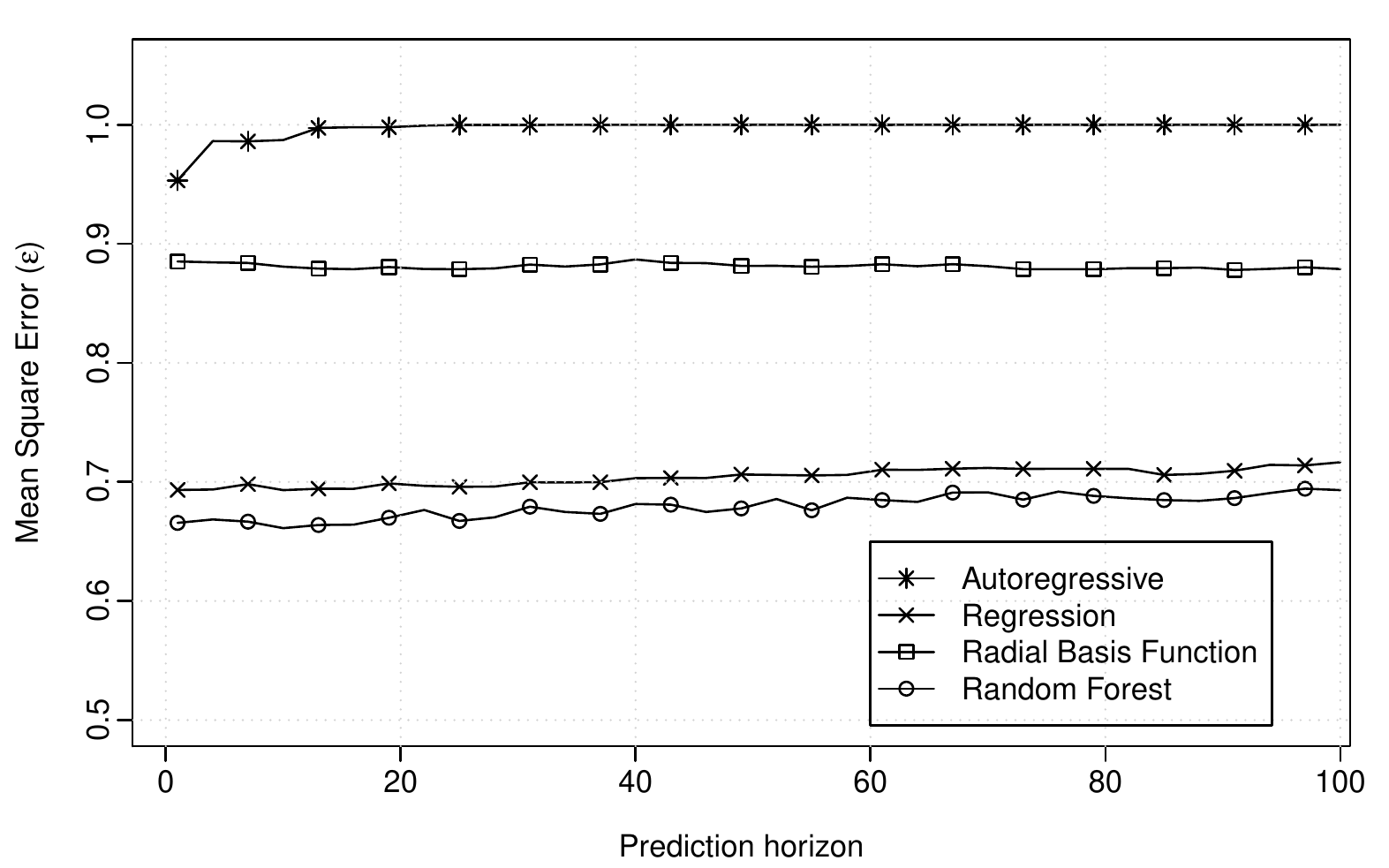}
\caption{\label{fig:MSEComp1} Mean square error comparison for different predictive models (Data set 1).}
\vspace{5mm}
\centering
\includegraphics[width=0.8\textwidth,height=0.5\textwidth]{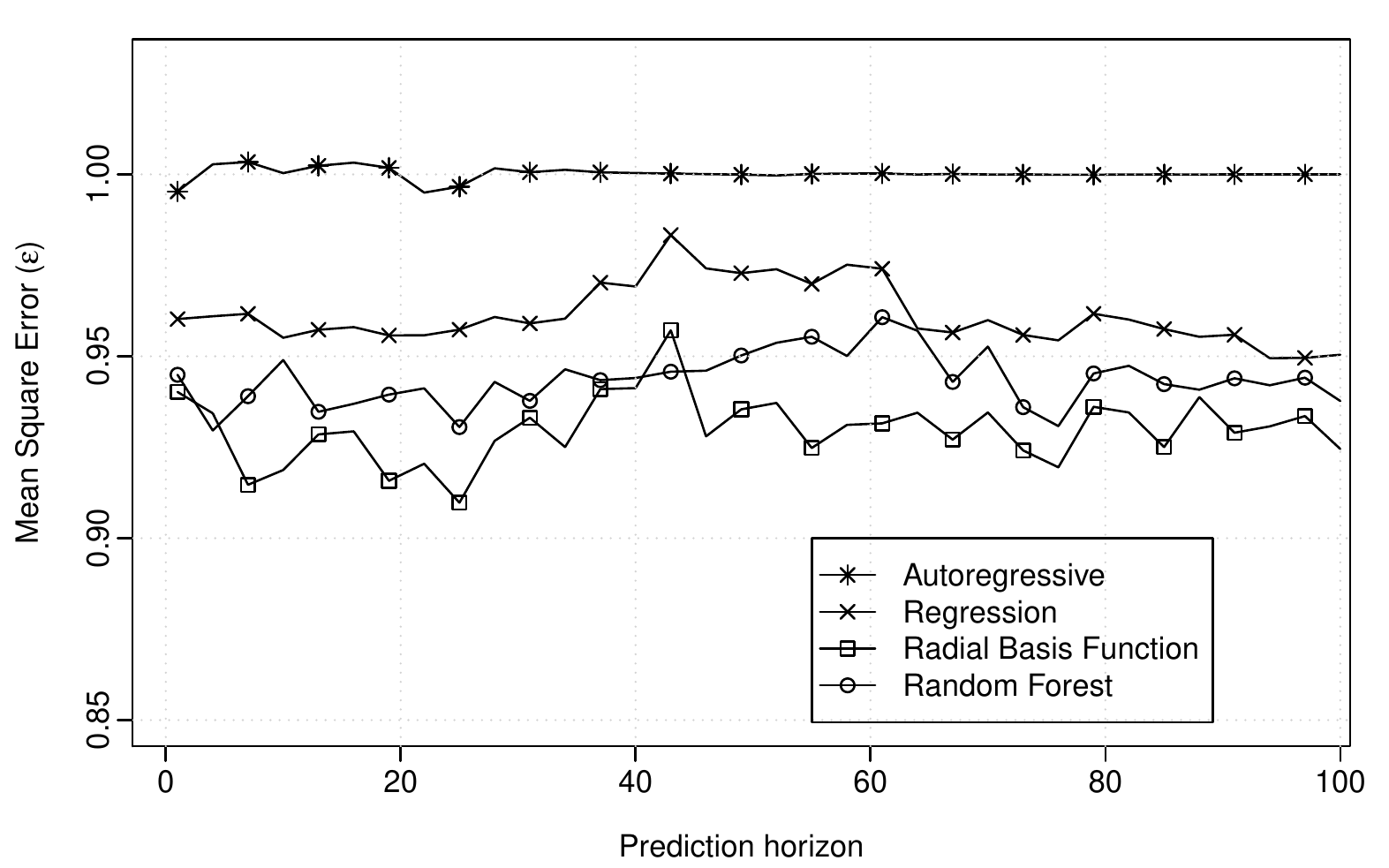}
\caption{\label{fig:MSEComp2} Mean square error comparison for different predictive models (Data set 2).}
\end{figure}

The relative performance for different prediction horizons using models from Tables \ref{tab:compCF1a} and \ref{tab:compCF2a} is plotted in Figures \ref{fig:MSEComp1} and \ref{fig:MSEComp2} respectively. As expected, autoregressive models performed no better than the mean as a naive forecaster. In the case of data set 1, random forests and regression models performed better than radial basis functions. On the other hand, a small difference in performance was found between regression, radial basis function and random forest models for data set 2.

Results for the rest of our illustration data sets are shown in Table \ref{tab:ModelPAComp}. In data set 1, a small difference in forecasting accuracy was found in favor of the random forest model in comparison to a regression model. Additionally, the performance of radial basis function models in the data set 1 was worse than the regression and random forest models but it was better in the data set 2. From that, it is clear that forecasting accuracy of the autoregressive model can be improved by considering regression, radial basis functions and random forest models using some basic explanatory variables. Next, we measure the savings produced by a better prediction.

\begin{table}[ht]
\caption{Average predictive accuracy for prediction horizons from 1 to 100 days using Method 1. Standard deviations are shown in parenthesis and best values are bold.}
\begin{center}
\begin{tabular}{ c p{2.2cm} p{2 cm} p{2cm} p{2.0cm}}
\hline
Data set & Autoregressive & Regression & Radial Basis Function & Random Forest\\ \hline 
1  & 0,998 & 0,704 & 0,880 & \textbf{0,680} \\
   &  (0,007) &  (0,007) &  (0,003) &  (0,009) \\
2  & 0,999 & 0,962  & \textbf{0,930}  & 0,942 \\
  &  (0,002) &  (0,008) &  (0,008) &  (0,006) \\
3  & 0,997  & 0,697 & 0.870) & \textbf{0,669} \\
  &  (0,008) &  (0,007) & (0.004) &  (0,009) \\
4  & 0,998 & 0,763  & 0.952  & \textbf{0,749}  \\
  &  (0,006) &  (0,007) & (0.002) &  (0,010) \\
5  & 0,999  & 0,826  & 0.902 ) & \textbf{0,821}  \\
  & (0,003) & (0,004) & (0.002) & (0,006) \\
6  & 0,999  & 0,977  & \textbf{0.946}  & 0,949  \\
  & (0,003) & (0,007) & (0.011) & (0,009) \\
\hline
\end{tabular}
\end{center}
\label{tab:ModelPAComp}
\end{table}

\section{Does a better forecast produce a better cash-management policy? \label{section:PredAcComp}}

\citet{gormley2007utility} proposed a Dynamic Simple Policy (DSP) to demonstrate the utility of cash flow forecasts in the management of corporate cash balances. They proposed the use of an autoregressive model as the main input to their model. However, gains in forecast accuracy over a naive mean model were scant. Gormley and Meade expected that savings obtained using a non-naive forecasting model would increase if there were more systematic variation in the cash flow and, consequently, higher forecast accuracy. In the previous section, we showed that a better cash flow prediction can be obtained by using different forecast models. In this section, we verify that a better prediction produces a better policy. As a consequence, we find that the savings produced by a better forecasting model are significantly higher than those obtained by a naive forecasting model. 

Here, we exploit a simple policy equivalent to that of Gormley and Meade using the best forecasters as detailed in Section \ref{section:Evaluation} and compare to the results obtained by a constant mean forecast. Since the forecast accuracy of the autoregressive model almost equals the mean forecast accuracy (Tables \ref{tab:compCF1a} and \ref{tab:compCF2a}), the comparison to the mean is equivalent to the comparison to the autoregressive model. In what follows, we firstly introduce our empirical settings; secondly, we show empirically that forecasting accuracy leads to cost savings in the corporate cash management problem using a simple policy; and finally, we analyze potential cost savings of improving predictive accuracy of daily cash flow forecasts and a simple policy.

\subsection{Empirical settings}

The corporate cash management problem can be approached from a stochastic point of view by allowing cash balances to wander between two limits: the lower ($D$) and the upper balance limit ($V$). When the cash balance reaches any of these limits a cash transfer is made to return to the corresponding rebalance level ($d,v$). A model of this kind using daily forecasts was proposed by \citet{gormley2007utility} as a trade-off between transaction and holding costs that can be summarized as follows:

\begin{itemize}
\item q: Holding cost per money unit of positive balances at the end of the day. 
\item u: Shortage cost per money unit of negative balances at the end of the day.
\item $\gamma_{0}^{+}$: Fixed cost of transfer into account.
\item $\gamma_{0}^{-}$: Fixed cost of transfer from account.
\item $\gamma_{1}^{+}$: Variable cost of transfer into account.
\item $\gamma_{1}^{-}$: Variable cost of transfer from account.
\end{itemize}

Recall from section \ref{section:data} that we are dealing with a real business problem. Hence, we focus on current costs charged by banks to industrial companies in Spain. Current bank practices tend to charge a fixed cost for transfers between \euro1 and \euro5 and no variable cost so that we set $\gamma_{1}^{+}=0$ and $\gamma_{1}^{-}=0$. The shortage cost (\emph{u}) per money unit of a negative cash balance is around 30\% which represents a high penalty for negative cash balances. Finally, the holding cost (\emph{q}) per money unit of a positive cash balance is an opportunity cost of returns not obtained from alternative investments. Since this is not an actual cost but an opportunity cost based on judgmental criteria we set a range between 10\% p.a.\footnote{Per annum.} and 20\% p.a. based on expected alternative investments returns. We firstly try 15 different cost structures considered as the most likely scenario under current costs in Spain, denoted by (1) in Table \ref{tab:CS}. We also consider two additional scenarios, denoted by (2) and (3), to evaluate the effect of changes in particular costs. The second scenario tests the variation of the shortage cost ($u$) and the third one considers the introduction of variable transfer costs ($\gamma_{1}$).

\begin{table}[h]
\caption{Cost scenarios.}
\begin{center}
\begin{tabular}{l p{2.1cm} p{1.9cm} p{2.4cm}} 
\hline
Cost & Most likely scenario (1) & \multicolumn{2}{c}{Alternative Scenarios} \\ 
\cline{3-4}
{}   & {}   & Varying shortage cost $u$ (2)   & Introduction of variable cost $\gamma_{1}$ (3) \\ \hline 
Holding cost $q$ & 10, 15, 20\% & 15 \% p.a. & 15 \% p.a. \\ 
Shortage cost $u$ & 30 \% & 10, 20, 40 \% &  30 \% \\
Fixed into account $\gamma_{0}^{+}$ & 1, 2, 3, 4, 5 \euro & 3 \euro & 3 \euro \\
Fixed from account$\gamma_{0}^{-}$ & 1, 2, 3, 4, 5 \euro & 3 \euro  & 3 \euro\\
Variable into account$\gamma_{1}^{+}$ & 0 \% & 0 \% & 0.1, 0.2, 0.4 \permil \\
Variable from account $\gamma_{1}^{-}$ & 0 \% & 0 \% & 0.1, 0.2, 0.4 \permil \\
\hline
\end{tabular}
\end{center}
\label{tab:CS}
\end{table}

In our experiments, parameter selection of the cash management model is performed under a business perspective. In \citet{gormley2007utility} the policy parameter values \emph{D, d, v, V} were chosen to minimize the expected cost over horizon \emph{T} using a genetic algorithm \citep{chelouah2000continuous}. Here, since the focus is placed on the comparison between policies obtained from different forecasting models, parameter optimization plays a secondary role. Therefore, these parameters are empirically chosen and kept unaltered in the comparison between savings for each forecasting model and each cost scenario. However, in order to evaluate the influence of these parameters on the utility of the forecast, three different cases are studied based on risk tolerance. Since the cost of a negative balance is very high, common sense leads us to set \emph{D} to a minimum level so that only a given percentage (\emph{MaxPct}) of expected cash flows can bring the balance from value \emph{D} to a negative value. The higher the percentage, the higher the probability of an overdraft and, consequently, the riskier the policy under these cost structures. We study three cases with different levels of risk:

\begin{enumerate}
\item Low risk or $MaxPct=5\%$. 
\item Medium risk or $MaxPct=10\%$.
\item High risk or $MaxPct=15\%$. 
\end{enumerate}

On the other hand, the use of dynamic simple policy assumes that an unlimited cash buffer is available to transfer into the bank account whenever it is necessary. In practice, this situation is unrealistic. Thus, we restrict high balance levels by setting an upper limit to 1.5 times the lower cash balance limit. Following the recommendations in \citet{gormley2007utility}, the positive shift from the lower balance limit ($D$) of lower rebalance level ($d$) is proportional to the difference between the higher ($V$) and the lower balance ($D$) limits. Finally, the negative shift from the higher balance limit ($V$) to obtain the higher rebalance level is proportional to the difference between the higher balance limit ($V$) and the lower rebalance level ($d$). Here, we chose proportionality constants $\alpha_1=0.5$ and $\alpha_2=0.5$ to produce an even distance between policy parameters. The entire analysis would remain the same when varying this setting. As a summary, parameters selection is done according to:

\begin{itemize}
\item $D=|o_{th}|$ where $o_{th}$ is the $N \cdot MaxPct$-th element of vector $o_t$ of ascending ordered values of cash flow being $N$ the total number of observations. 
\item $V=1.5D$, then $V-D=\frac{D}{2}$ 
\item $d=D+\alpha_1(V-D)$ with $\alpha_1=0.5$
\item $v=V-\alpha_2(V-d)$ with $\alpha_2=0.5$.
\end{itemize}

Predicted cash flows using different forecasters are used to compare the effect on the total cost over different prediction horizons ($h$) from 1 up to 100 days ahead. We set $g$ to the minimum number of observations required to estimate the model that is equivalent to 65\% of the data set. We proceed as detailed in Algorithm \ref{Alg2}.
\\

\begin{algorithm}[H]
 \caption{Comparison algorithm \label{Alg2}}
 \textbf{Input}:Cash flow data set of \emph{T} observations, $g, h, MaxPct$, and a forecaster\;
 \textbf{Output}:Average cost difference between a forecaster and the mean as a forecast\;
 \For{$i=1,2,\ldots,T-g-h+1$}
  {
  Estimate the model with observations at times $1,2,\ldots,g+i-1$\;
  Predict for times $g+i$ up to $g+h+i$ using the forecaster\;
  Predict for times $g+i$ up to $g+h+i$ using the mean forecaster\;
  	\For{$j=1,2,\ldots$, Number of cost structures}
    {
    	Compute cost for the \emph{i}th forecast when using the \emph{j}th structure\;
        Compute cost for the \emph{i}th mean forecast and the \emph{j}th structure\;
    }
  }
 Compute average cost for each cost structure using the forecaster\;
 Compute average cost for each cost structure using the mean forecaster\;
 Compute difference between average cost of the mean and the forecaster\;
\end{algorithm}
\vspace{5mm}

\subsection{Impact of predictive accuracy on cost savings}

Cost savings are computed as the daily average cost differences between the naive forecast and the best-in-class forecaster for each of the data sets (Table \ref{tab:Cost}). Recall that this comparison to the mean is equivalent to the comparison to the autoregressive model. 

From these results, we can say that, in general, an increase in forecast accuracy leads to significant cost savings using a simple policy. A better forecasting model produces higher savings for either conservative or riskier policies (Table \ref{tab:Cost}). The effect of forecasting accuracy in daily costs dramatically rises when the policy parameters are reduced as a consequence of a riskier policy. In these cases, forecasting accuracy is much more important in reducing daily cost due to the risk of an overdraft. As expected, cost reductions for the data set 2 are smaller but still significant due to less predictive accuracy.

A deeper insight on the different scenarios shows that changes in cost parameters have a reduced impact on cost savings. Moreover, changes in the variability of cash flow data, studied here by introducing less (data set 3) or more variance (data set 4), produced no major changes. However the effect of random shocks in the cash flow data (data sets 5 and 6), reduced cost savings due to the higher uncertainty of the cash flow data.

\begin{table}[h]
\caption{Average daily saving for different levels of risk and the most likely scenario. RF=Random Forest; RBF=Radial Basis Function; $u=$ shortage cost, $\gamma_1=$ variable transaction cost.}
\begin{center}
\begin{tabular}{p{0.6cm} p{1.2cm} l r r r}
\hline 
Data set & Best-in-class & Scenario & Low Risk & Medium Risk & High Risk \\ \hline 
	1 & RF & Most likely & 182 (64\%) & 1398 (72\%) & 2034 (54\%) \\
    1 & RF & Varying $u$    & 141 (54\%) & 1088 (71\%) & 1583 (54\%) \\
    1 & RF & Introducing $\gamma_{1}$    & 183 (61\%) & 1399 (72\%) & 2035 (54\%) \\
    2 & RBF & Most likely    & 135 (51\%) & 479 (52\%) & 686 (44\%) \\
    2 & RBF & Varying $u$     & 105 (47\%) & 372 (51\%) & 534 (44\%) \\
    2 & RBF & Introducing $\gamma_{1}$   & 135 (49\%) & 479 (51\%) & 687 (44\%) \\
    3 & RF & Most likely    & 180 (64\%) & 1386 (72\%) & 2017 (54\%) \\
    4 & RF & Most likely    & 182 (63\%) & 1412 (73\%) & 2083 (52\%) \\
    5 & RF & Most likely    & 58 (8\%) & 1219 (45\%) & 1798 (38\%) \\
    6 & RBF & Most likely    & 71 (15\%) & 355 (30\%) & 545 (28\%) \\
\hline
\end{tabular}
\end{center}
\label{tab:Cost}
\end{table}

\subsection{What if we improve predictive accuracy? Analyzing potential cost savings}

Our best-in-class forecasting models, i.e., radial basis functions and random forests, are attempts to reduce uncertainty in predicting daily cash flow. They represent two special cases in which improving predictive accuracy resulted in increasing cost savings over a naive forecast. However, cash managers may be interested in determining how much savings can be achieved by any extra effort in improving predictive accuracy. Since enhancing any forecasting model has a cost in terms of both time and money, it is important to know if this cost is offset by the savings obtained using a better forecasting. We can estimate savings associated to predictive accuracy by obtaining a number of synthetic predictions and evaluate the corresponding policy costs.

\citet{daellenbach1974cash} and \citet{da2014evolutionary} synthesized cash flow data for simulation purposes from normal distributions. Here, from a given cash flow time series ($y_{t+h}$), a new time series ($\hat{y}_{t+h}$) is synthesized by adding a random normal term of mean zero and a variable standard deviation ($\sigma$) using the following equation:

\begin{equation}
\hat{y}_{t+h}=y_{t+h}+\mathcal{N}(0,\sigma).
\label{eq:random}
\end{equation}

Increasing the value of $\sigma$, a set of time series with a decreasing degree of similarity to the original time series can be obtained. This is equivalent to generating a set of synthetic predictions with controlled predictive accuracy that can be evaluated in terms of mean square error ratio $\varepsilon(h)$ for different prediction horizons using equation (\ref{eq:MSE}).

For illustration purposes, here we obtain synthetic predictions covering a range from $\overline{\varepsilon}=0$ to values greater than 1. Here $\overline{\varepsilon}$ denotes the average of $\varepsilon(h)$ for prediction horizons up to 100 days on a test set formed by the last 35\% of the observations of data sets 1 and 2. Later, savings for each of these synthetic forecasts are obtained following Algorithm \ref{Alg2} but using the synthetic forecasts previously generated rather than estimating and predicting.

\begin{figure}[!htb]
\centering
\includegraphics[width=0.8\textwidth,height=0.5\textwidth]{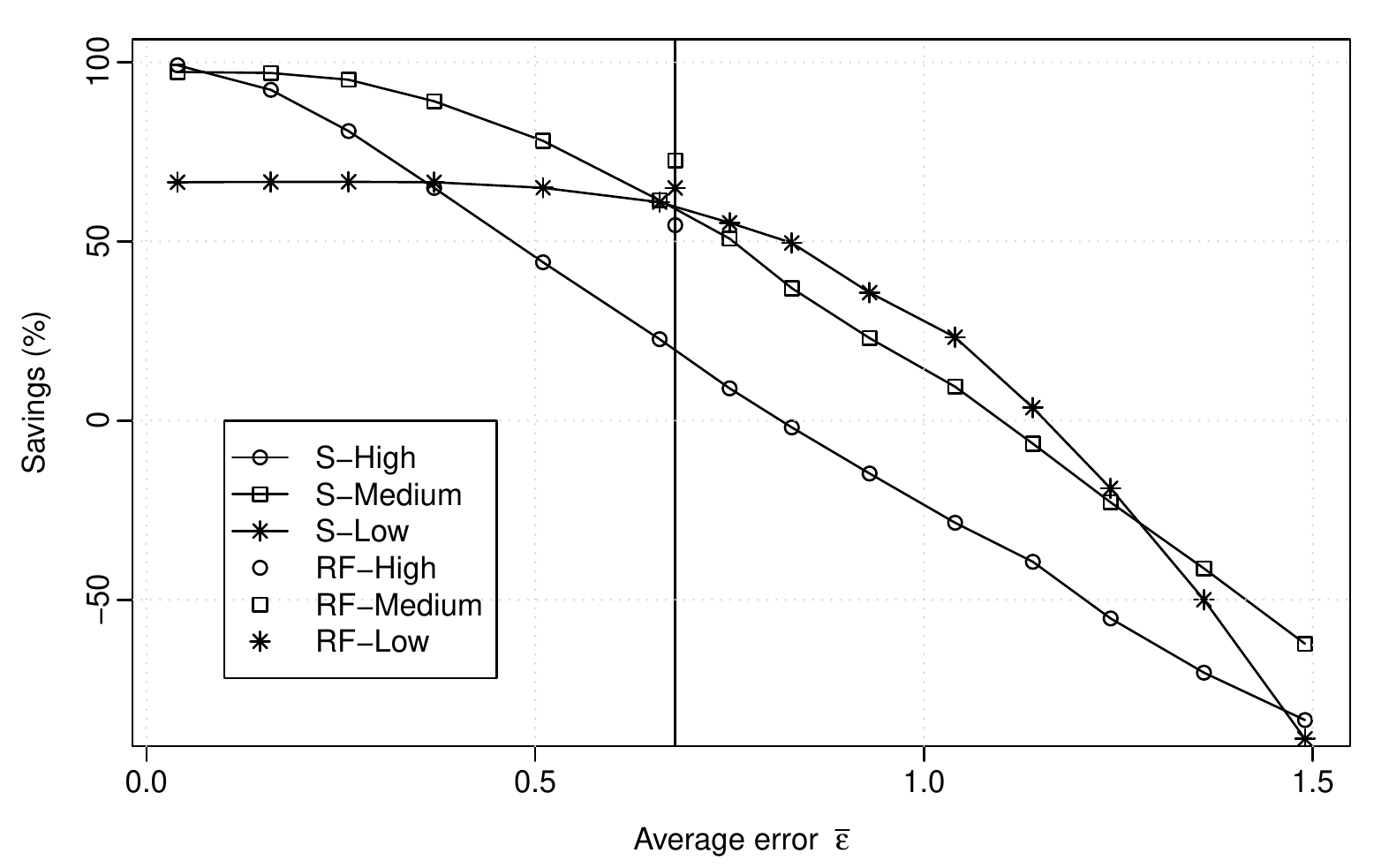}
\caption{\label{fig:PAComp1} Savings for different predictive errors and levels of risk for data set 1 in the most likely scenario. S=Synthetic forecasts, RF=Random Forest.}
\vspace{5mm}
\centering
\includegraphics[width=0.8\textwidth,height=0.5\textwidth]{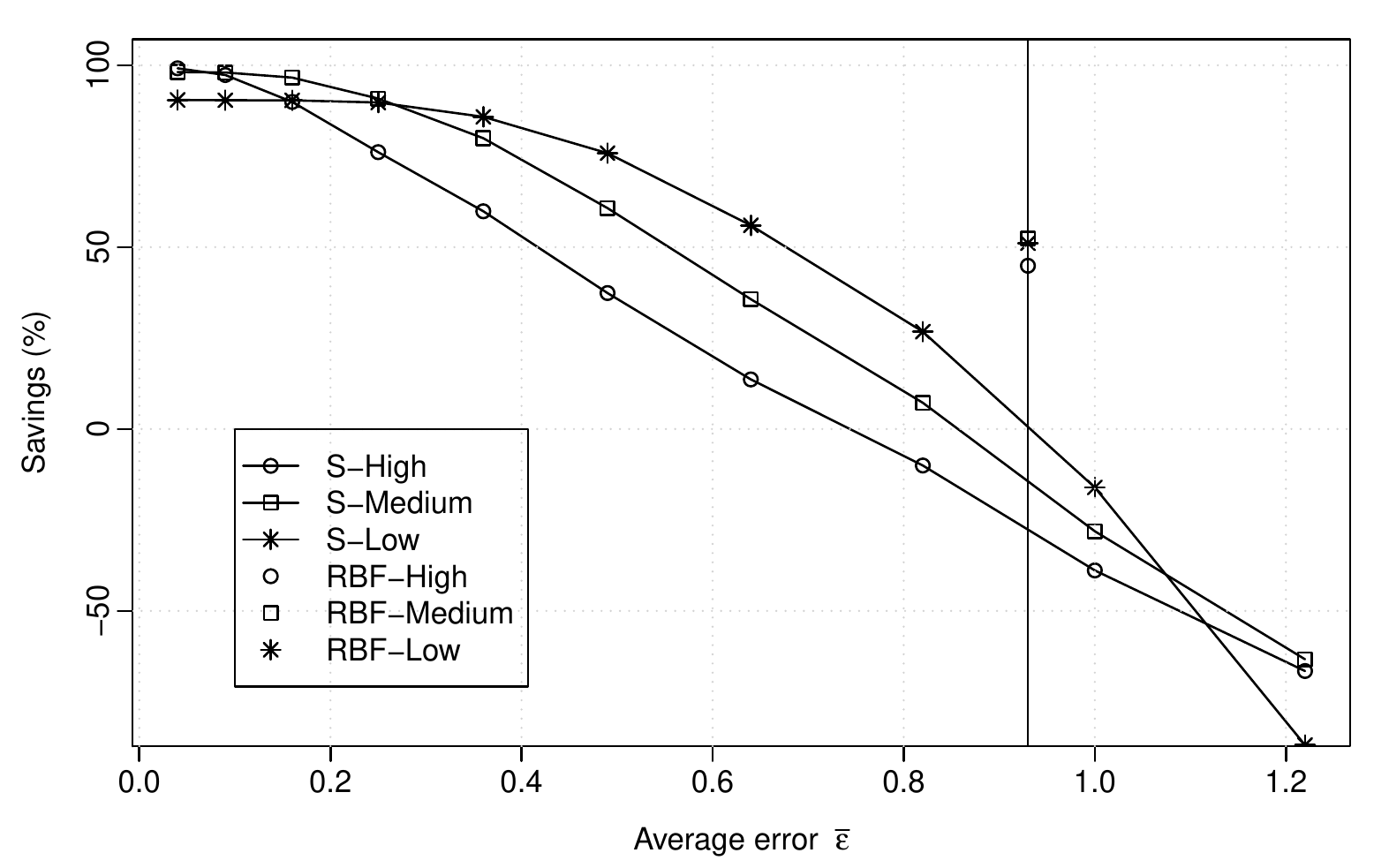}
\caption{\label{fig:PAComp2} Savings for different predictive errors and levels of risk for data set 2 in the most likely scenario. S=Synthetic forecasts, RBF=Radial Basis Function.}
\end{figure}

Results from this simulation for data sets 1 and 2 and three different levels of risk are shown in Figures \ref{fig:PAComp1} and \ref{fig:PAComp2}. As a reference, the vertical lines locate savings achieved by the best-in-class forecaster for each of the examined levels of risk. For example, using random forests for data Set 1, a value of $\overline{\varepsilon}=0.68$ (from Table \ref{tab:compCF1a}) was obtained which produced savings of 54, 72 and 64\% (from Table \ref{tab:Cost}) for the three levels of risk considered.

As expected, improving prediction accuracy, i.e., reducing $\overline{\varepsilon}$, leads to an important increase in cost savings up to 100\% in the case of a perfect prediction. Efforts in increasing predictive accuracy are notably rewarded. However, the behavior is different depending on the level of risk chosen by the company.

\begin{enumerate}
\item Low risk: the effect of improving predictive accuracy tends to a stable point where any further effort yields no additional saving. In spite of the considerable percentage saved, it seems that improvement potential in predictive accuracy is limited when the risk is low.
\item Medium risk: the effect of limited cost savings when improving predictive accuracy is also present but to a lesser extent. 
\item High risk: the behavior is almost linear in the considered $\overline{\varepsilon}$ interval.
\end{enumerate}

It is interesting to point out that the relationship is almost linear in most of the range of the average error $\overline{\varepsilon}$ for each of the three levels of risk. This fact should encourage practitioners to work hard to obtain a better prediction because they can expect a proportional reward in terms of cost savings. However, in the case of our best-in-class forecasting model using random forests and data set 1, an error $\overline{\varepsilon}$ of 0.68 (from Table \ref{tab:compCF1a}) places the savings in the highest value likely to be obtained for the low level of risk. Any effort in improving predictive accuracy will be useless. This behavior is confirmed by the fact that a perfect prediction was unable to achieve a 100\% difference in cost savings.

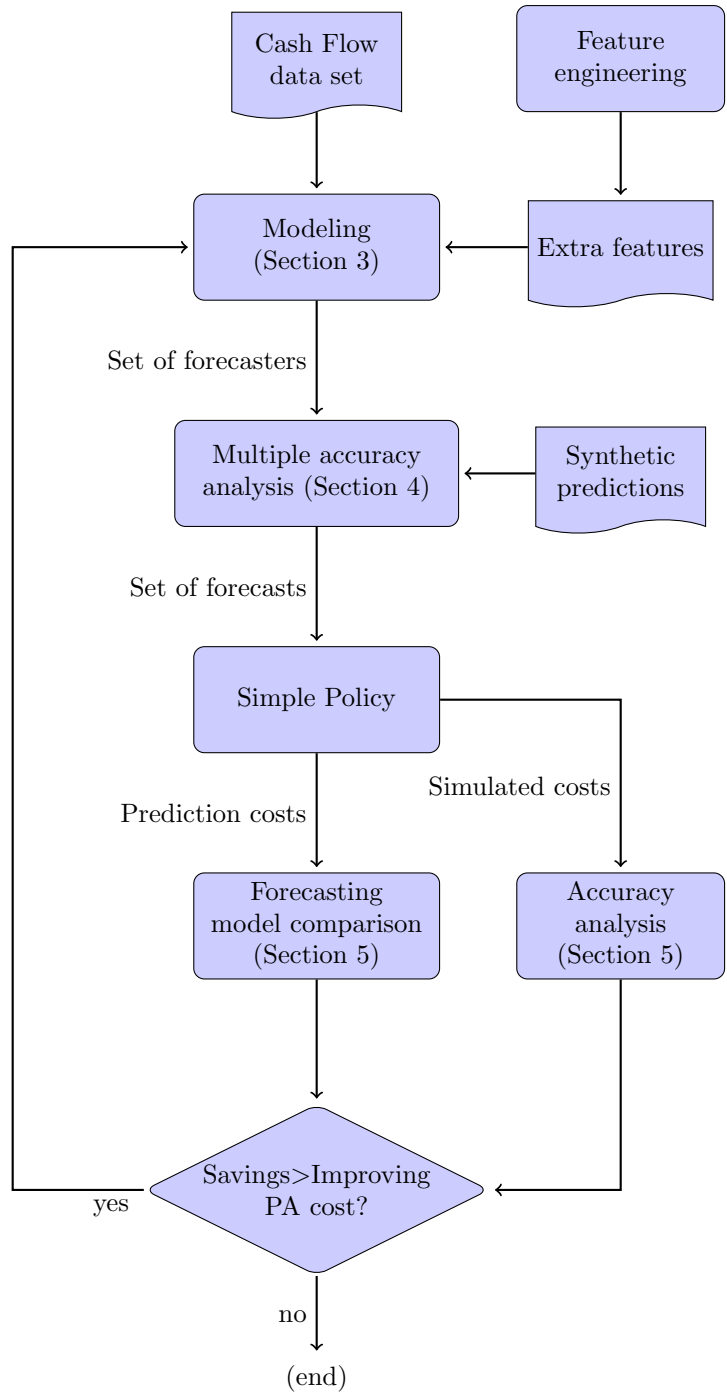
\begin{figure}[]
\centering
\begin{tikzpicture}[node distance = 3cm, auto]

    \node [print, text width=2cm, node distance=5.5cm] (data2) {Cash Flow data set};
    \node [block, below of=data2, node distance= 2.5cm] (Modeling) {Modeling (Section \ref{section:models})};
    \node [block, below of=Modeling, text width=3.5cm] (AccuracyComp2) {Multiple accuracy analysis (Section \ref{section:Evaluation})};
	\node [block, below of=AccuracyComp2] (DSP2) {Simple Policy};
    \node [block, below of=DSP2] (Comp2) {Forecasting model comparison (Section \ref{section:PredAcComp})};
    \node [block, right of=data2, node distance= 4cm, text width=2.5cm] (FeatureEng) {Feature engineering};
    \node [print, below of=FeatureEng, node distance=2.5cm] (data3) {Extra features};
    \node [block, right of=Comp2, text width=2.5cm, node distance=4cm] (PredictiveComp) {Accuracy analysis (Section \ref{section:PredAcComp})};
	\node [print, right of=AccuracyComp2, node distance=4cm, text width=2cm] (Preds) {Synthetic predictions}; 
    \node [decision, below of=Comp2, node distance=3.5cm, text width={3cm}] (FurtherInvest) {Savings$>$Improving PA cost?};
    \node [text centered, below of=FurtherInvest, node distance=2.5cm] (end) {(end)};  
 
    \path [line] (data2) -- node [anchor=east] {} (Modeling);
    \path [line] (Modeling) -- node [anchor=east] {Set of forecasters} (AccuracyComp2);
    \path [line] (FeatureEng) -- node [anchor=east] {} (data3);
	\path [line] (data3) -- node [anchor=south] {} (Modeling);
	\path [line] (AccuracyComp2) -- node [anchor=east] {Set of forecasts} (DSP2);
    \path [line] (DSP2) -- node [anchor=east] {Prediction costs} (Comp2);
    \path [line] (Preds) -- node [anchor=south] {} (AccuracyComp2);
    \path [line] (DSP2) -| node [near end, anchor=east] {Simulated costs} (PredictiveComp);
    \path [line] (Comp2) -- (FurtherInvest); 
    \path [line] (PredictiveComp) |- (FurtherInvest);
    \path [line] (FurtherInvest) --++ (-4,0) node [near start] {yes} |- (Modeling); 
    \path [line] (FurtherInvest) -- node [anchor=east] {no} (end);
\end{tikzpicture}
\caption{\label{fig:Methodology}Potential cost saving analysis methodology (PA=Predictive Accuracy)}
\end{figure}

Summarizing, we propose a new and more comprehensive methodology for cash managers, as shown in Figure \ref{fig:Methodology}, based on the effect of predictive accuracy on cash management cost using daily cash flow forecasts and a simple policy. In order to allow different models to capture patterns, cash managers should consider an additional previous step of feature engineering to obtain a series of extra features. They can also adopt a wider modeling approach that allow them to compare a set of forecasters in terms of forecasting accuracy. At this point, cash managers can easily generate a number of synthetic predictions to cover a wide range of different predictive accuracy by tuning a parameter. These synthetic predictions, and those obtained using our best-in-class forecasters from Table \ref{tab:ModelPAComp}, are tested in their ability to reduce the cost of the policies by using a simple policy. This step results in a graphical estimation on how much cost savings can be achieved by improving predictive accuracy of our selected forecasters. If estimated savings are greater than the cost of improving the accuracy of the forecasting models, a new modeling process is worth undertaking.

\section{Conclusions and future work \label{sec:conclusion}}

From the above-described results, we derive two main findings. First, assessing predictive accuracy is a must in the context of corporate cash management, specially when employing daily forecasts as an input to a cash flow management model. Indeed, we empirically find that cost savings are highly sensitive to improvements on prediction accuracy when using a simple policy, and hence major savings may stem from accurate predictions. Second, from a cost sensitive perspective, cash managers may consider our methodology to decide whether improving the predictive accuracy at hand is financially worthy. These two findings, which we further dissect next, are meant to yield benefits for cash managers.
\\

\noindent
\textbf{On the impact of predictive accuracy on cost savings.} \citet{gormley2007utility} hypothesized that the more accurate the cash flow forecasting accuracy, the larger the cost savings expected. Here, for the first time, we have empirically confirmed such hypothesis. Furthermore, we have analyzed the impact of predictive accuracy on average daily cost savings when considering a variety of cost structures (of real-world bank finance conditions) and cash flow policy parameters. From our analysis we have learned that:

\begin{itemize}
\item \emph{Predictive accuracy is strongly correlated with cost savings} when using daily forecasts in cash management models. Thus, cost savings were highly sensitive to improvements on prediction accuracy when using a simple policy and two real-world cash flow data sets.
\item \emph{The riskier the cash management policy, the higher the average daily cost reduction in cash.}
\item \emph{The realistic cost structures considered in the most likely scenario have little influence on cost savings obtained by the forecasting models.} 
\end{itemize}

\noindent
\textbf{What if predictive accuracy increases? Analyzing potential savings.} 
Cash managers may wonder if efforts on improving forecasting accuracy are expected to be proportionally rewarded by cost savings. Along this direction, we proposed a method for estimating the cost savings potentially delivered by improving predictive accuracy. Independently of the predictive accuracy of the forecaster available to a cash manager, our results help her estimate the cost savings that she might expect. Moreover, even if the cash manager does not count on any forecaster, she can estimate the cost savings that she currently misses. Overall, we learned that different risk levels yield different estimation results so that:

\begin{itemize}
\item \emph{When assuming low risk, cost savings are limited} and further efforts in enhancing predictive accuracy are expected to be useless, in terms of both time and money, when a particular point in predictive accuracy is reached; and
\item \emph{The higher the risk a cash manager assumes, the higher the expected reward when improving predictive accuracy}.
\end{itemize}

The analysis of the relationship between predictive accuracy and cost savings leads to confirm the importance of better forecasting models when predictions are used as the main input to cash management models. Some additional intuition can be derived in the sense that this behavior may be caused by a number of reasons: (i) whenever it is possible to reduce uncertainty about the future, better decisions can be made; (ii) improving predictive accuracy is closely linked to discover patterns and an appropriate response to these patterns is necessarily useful; (iii) chances are that cash management models using forecasts as the main input do not work well with low quality forecasts. All of them highlight again the utility of forecasts in cash management.

To end up, besides the above-mentioned benefits from the cash manager perspective, the task of building forecasting models helped us also learn that daily seasonality influenced forecasting. However, future work is in place to search for a more informative set of features in the corporate cash management problem. In this sense, feature engineering is meant to play a key role to help improve predictive accuracy and ultimately produce cost savings in cash management.

\end{document}